\documentclass[apjl,twocolumn]{openjournal}

\usepackage{graphicx}
\usepackage{epstopdf}
\usepackage{url}

\begin{document}

\title{Spatial segregation of massive clusters in a simulation of colliding dwarf galaxies}
\author{Bruce G. Elmegreen\altaffilmark{1},
Natalia Lah\'en\altaffilmark{2}}
\altaffiltext{1}{Katonah, New York USA, belmegreen@gmail.com}
\altaffiltext{2}{Max-Planck-Institute f\"ur Astrophysik, Karl-Schwarzschild-Straße 1, D-85740 Garching, Germany}

\begin{abstract}
The collective properties of star clusters are investigated using a simulation of the collision between two dwarf galaxies. The characteristic power law of the cluster mass function, $N(M)$, with a logarithmic slope $d\log N/d\log M\sim-1$, is present from cluster birth and remains throughout the simulation. The maximum mass of a young cluster scales with the star formation rate (SFR). The relative average minimum separation, ${\cal R}(M)= N(M)^{1/p}{\bar D}_{\rm min}(M)/{\bar D}(M_{\rm low})$, for average minimum distance ${\bar D}_{\rm min}(M)$ between clusters of mass $M$, and for lowest mass, $M_{\rm low}$, measured in projection ($p=2$) or three dimensions ($p=3$), has a negative slope, $d\log{\cal R}/d\log M\sim -0.2$, for all masses and ages. This agrees with observations of ${\cal R}(M)$ in low-mass galaxies studied previously.  Like the slope of $N(M)$, ${\cal R}(M)$ is apparently a property of cluster birth for dwarf galaxies that does not depend on SFR or time. The negative slope for ${\cal R}(M)$ implies that  massive clusters are  more concentrated relative to lower mass clusters throughout the entire mass range.   Cluster growth through coalescence is also investigated. The ratio of the kinetic to potential energy of all near-neighbor clusters is generally large, but a tail of low values in the distribution of this ratio suggests that a fraction of the clusters merge, $\sim8$\% by number throughout the $\sim300$ Myr of the simulation and up to 60\% by mass for young clusters in their first 10 Myr, scaling with the SFR above a certain threshold.
\end{abstract}

\section{Introduction} \label{sec-intro}
Stars form in dense interstellar gas that has a hierarchical structure from turbulence and self-gravity \citep[e.g.,][]{scalo85,stan99,elmegreen01}. Young stellar positions are similarly hierarchical with star clusters forming in the mixed inner parts of the densest regions \citep[e.g.,][]{zhang01,elmegreen10,guszejnov22}.  For a self-similar hierarchy, the mass distribution function of concentrations has a power law with a slope of around $-2$ \citep[$-1$ on a log-log plot;][]{fleck96,elmegreen96}, which explains the power law portion of the mass function that is commonly observed for clusters \citep{elmegreenefremov97,zhang99,krumholz19,cook19}.  The power-law portion of the stellar mass function may have a similar origin \citep{elmegreen97,hennebelle08,shadmehri11}. 

Power laws from structure rather than self-regulated processes have the property that the maximum mass of an object that forms scales with the total number of objects \citep{elmegreen83}. Such scaling has been observed for both star clusters \citep{larsen02,billett02,whitmore02,hunter03,bastian08,cook23} and stars \citep{corbelli09,andrews14,jung23}. Physical models for the dependence of  maximum cluster mass on star formation rate are in \cite{weidner04} and \cite{rand13}.

Another property of hierarchical structure is mass segregation, as shown by \cite{elmegreen99} for stars randomly selected from a hierarchical mass distribution.  Stellar mass segregation inside star clusters may also arise over time from stellar interactions \citep{spitzer40}, but its appearance in even the youngest clusters \citep{hillenbrand98,bonnell98,gennaro11,elmegreen14} suggests a primordial origin connected with the gas distribution.   Such stellar mass segregation caused us to wonder whether whole star clusters are also mass segregated relative to their neighboring clusters. Using 14 galaxies in the LEGUS survey, and several tidal dwarf galaxies in the \cite{fensch19} survey, we found that massive clusters are more centrally concentrated relative to other clusters in the dwarf galaxies but not in the spirals \citep{elmegreen20}.  Numerous other galaxy observations report radially varying cluster mass \citep[e.g.,][]{adamo15,rand19}. 

Massive cluster formation is common in dwarf galaxy collisions \citep{adamo20a,kimbro21,micic23}.  The purpose of the present paper is to look for cluster mass segregation in a solar-mass resolution simulation of two colliding dwarf galaxies \citep[Sect. \ref{sims};][]{lahen19, lahen20}. We use the method of our previous study, which involves the mass-derivative of a dimensionless function equal to the ratio of the average minimum separation between clusters of a particular mass and the expected average for a uniform density.  Section \ref{massfunction} discusses the cluster mass function, Section \ref{rampsection} the segregation and radial dependence of cluster mass, and Section \ref{coalesce} the possibility that some clusters coalesce. Some implications of hierarchical assembly and radial segregation are in Section \ref{discussion}, while the conclusions are in Section \ref{conclusions}.

\begin{figure}
\includegraphics[width=\columnwidth]{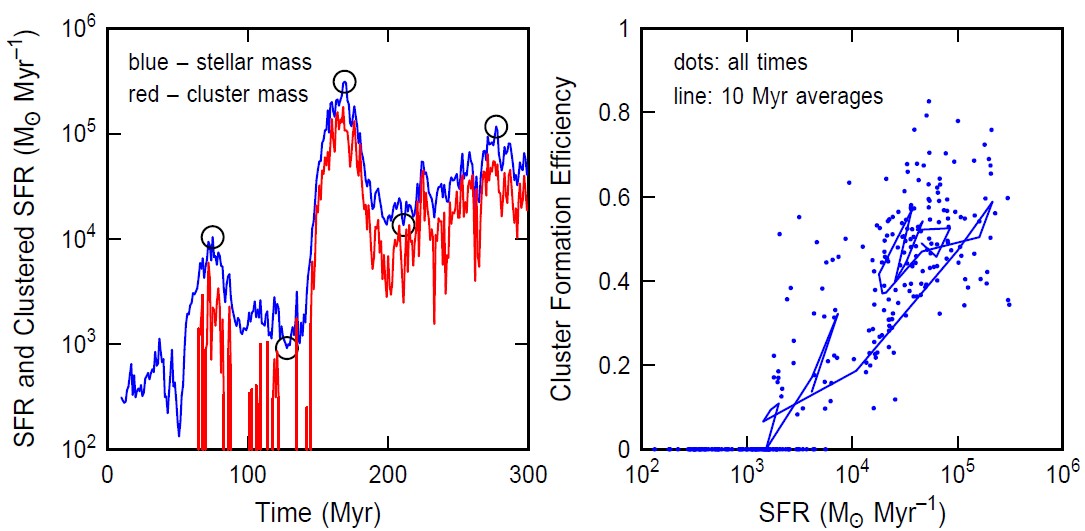}
\caption{Left: The star formation rate (blue, measured as the mass of new stars formed in Myr intervals) and the clustered star formation rate (red, total cluster mass formed per Myr) are shown as functions of time in the simulation. Five black circles indicate star formation rates at five fiducial times used in other figures below. Right: The cluster formation efficiency (CFE), which is the ratio of the CFR to the SFR in each Myr interval, has been averaged over 10 Myr intervals and plotted as a function of the SFR, which was also averaged over 10 Myr intervals, with straight lines connecting the sequence of time. The instantaneous CFE and SFR are shown as points. The CFE increases with the SFR beyond a SFR of $\sim10^3\;M_\odot$ (which may be an artificial limit as clusters with less than $\sim200\;M_\odot$ were not cataloged).}
\label{sfr}
\centering
\end{figure}

\section{Simulation}
\label{sims}
The cluster data used here comes from the hydrodynamical simulation project \textsc{griffin} (Galaxy Realizations Including Feedback From INdividual massive stars) introduced in \cite{lahen19} and \cite{lahen20}. The simulation models star and star cluster formation throughout a low-metallicity ($Z\sim0.1$ Z$_\odot$) starburst, occurring as two identical gas-rich dwarf galaxies interact and merge. The initial galaxy models have virial masses of $2\times 10^{10}$ M$_\odot$, including a $2\times 10^{7}$ M$_\odot$ stellar disk and a $4\times 10^{7}$ M$_\odot$ gaseous disk initially set in hydrostatic equilibrium following \citet{springel05a}. Gas mass resolution is 4 M$_\odot$ and gravitational softening is 0.1 pc for gas and stars. When new stars form, the stellar initial mass function of \citet{kroupa01} is sampled and masses between 1--50 M$_\odot$ are stored, to be used for instance in stellar feedback. All stars more massive than 4 M$_\odot$ are treated as individual stellar particles and lower mass stars are grouped in particles with a minimum mass of 4 M$_\odot$.

The simulations were run with \textsc{sphgal} \citep[and references therein]{hu17} that is a modern smoothed particle hydrodynamics (SPH) code based on \textsc{gadget-3} \citep{springel05b} with significant improvements to the SPH method as outlined in \citet{hu14}. \textsc{sphgal} models non-equilibrium cooling and heating processes in the low-temperature interstellar medium using a chemical network \citep{nelson97, glover07, glover12}. Star formation occurs stochastically when the local Jeans-mass is between $0.5$--$8$ SPH kernel masses (100 neighbours $\sim400$ M$_\odot$) at 2\% efficiency per free-fall time. For gas where the Jeans-mass subceeds 0.5 SPH kernel masses, star formation is enforced and such gas is immediately turned into stars. Stellar feedback in the simulation, modelled according to the initial mass and metallicity of each star, includes the spatially varying far-ultraviolet radiation field (with dust and gas self-shielding), photoionizing radiation, core-collapse supernovae \citep{chieffi04} and asymptotic giant branch winds \citep{karakas10}. Stellar lifetimes were adopted from \citet{georgy13}.

The integrated SFR, repeated here in Figure \ref{sfr}, spans three orders of magnitude from a few $10^{-4}$ M$_\odot$ yr$^{-1}$ to a few $10^{-1}$ M$_\odot$ yr$^{-1}$. The evolving SFR provides a wide range of cluster-forming environments. Star clusters with gravitationally bound masses up to $\sim8\times 10^{5}$ M$_\odot$ form in hierarchically structured cluster-forming regions. Catalogues of gravitationally bound star clusters with at least 50 stellar particles ($\sim200$ M$_\odot$) were constructed in \citet{lahen20} using the structure finding algorithm \textsc{subfind} \citep{springel01} built in \textsc{gadget-3}. The cluster catalogues were recorded every 1 Myr. Here we utilize the mass, centre of mass, velocity and mass-weighted mean stellar age of the clusters.

\section{Cluster Mass Functions}
\label{massfunction}
Figure \ref{sfr} (left) shows the SFR and the clustered SFR (mass of clusters forming in each Myr interval, or cluster formation rate CFR) versus time for the simulation. Five fiducial points indicated by black circles occur at the main peaks and valleys. The selected epochs correspond to the first small burst of star formation in the tidal bridge after the first pericentric passage, the intermediate stage between the first and second passage, the peak of star formation after the second passage, the post-starburst period of relatively low activity, and a later time $\sim 100$ Myr after the starburst when star formation again picks up in the highly turbulent post-merger galaxy\footnote{See \url{https://wwwmpa.mpa-garching.mpg.de/~naab/griffin-project/movies.html} for a visualization}. 

\begin{figure}
\includegraphics[width=\columnwidth]{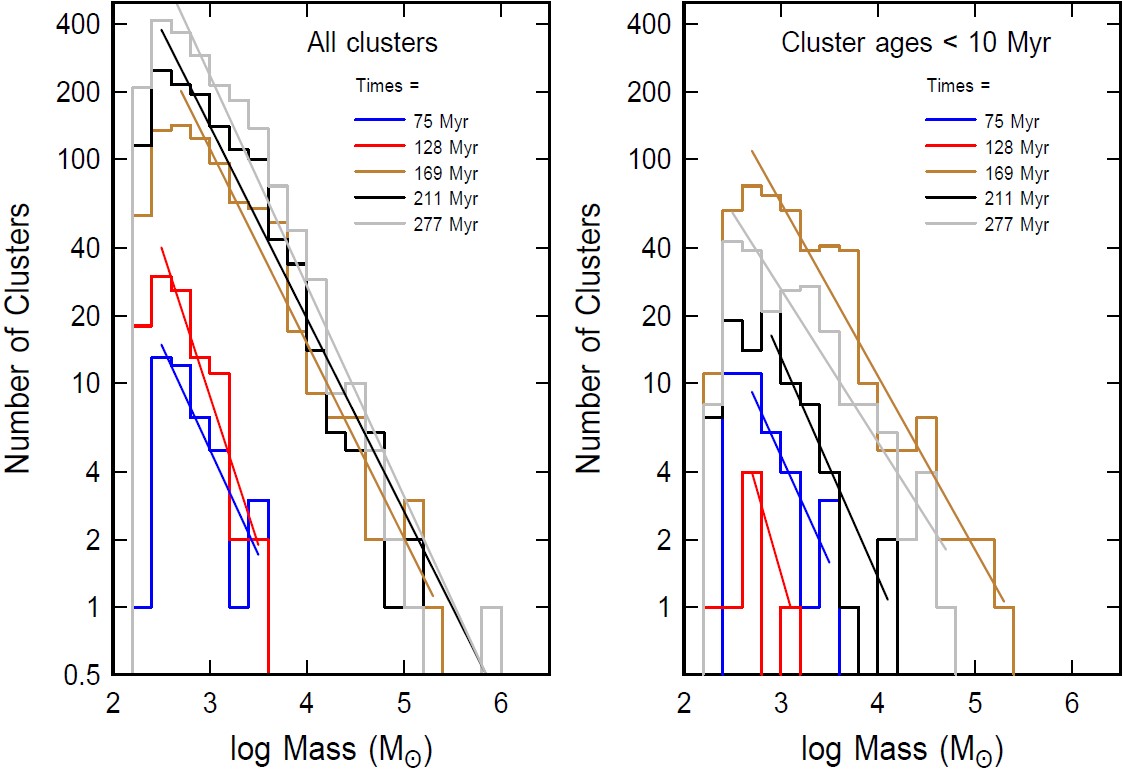}
\caption{The mass functions of clusters at the 5 fiducial times  in increasing order of the colors: blue, red, brown, black, gray. The mass functions for all clusters are on the left and for clusters younger than 10 Myr are on the right. Power-law fits to the declining parts of the functions are shown in the same colors. The logarithmic slopes are all around $-1$, as indicated in Table 1. }
\label{mf}
\centering
\end{figure}

The CFR generally traces the SFR, except when there is so little star formation that no massive enough ($\gtrsim 200$ M$_\odot$) star clusters form and the cluster formation rate is zero. The ratio of the CFR to the SFR, which is the cluster formation efficiency, CFE, increases with SFR in these simulations \citep[Figure \ref{sfr}(right); shown also in Fig. 8 of][]{lahen20}.  The line segments in this figure connect 10 Myr average values for both quantities and the dots show the values at each time, in Myr intervals. For the 10 Myr averages, this means the average of all the CFE ratios and all the SFRs in that 10 Myr, and not the total mass of clusters formed divided by the total mass of stars formed in the 10 Myr, which is closer to what is observed in real galaxies. We show the average of the CFE ratios because that is closer to the physical process of star formation, which works on 1 Myr timescales. Observations are ambiguous whether the ratio of the CFR to the SFR increases with SFR as clearly as in Figure \ref{sfr} \citep{chandar17,rand19,adamo20a,adamo20b,cook23,chandar23}.

Figure \ref{mf} plots the cluster mass functions at the five fiducial times in the following sequence of colors: blue for the first time, 78 Myr, followed by red (128 Myr), brown (169 Myr), black (211 Myr), and gray for the last time, 277 Myr. The mass functions for all the clusters regardless of age are on the left and for clusters younger than 10 Myr are on the right. The straight lines are linear fits (on these log-log plots) to the right of the peaks. The slopes $\alpha$ of these lines are the power law indices of the mass functions. They are tabulated in Table \ref{table:slopes}. For a mass function written in linear intervals of mass, the slope is 1 less than $\alpha$, i.e., $n(M) \propto M^{\alpha-1}$ for masses between $M$ and $M+dM$ (i.e., $dn(M)/dM\propto M^{\alpha-1}$ and $d\log N(M)/d\log M\propto M^{\alpha}$). The highest-mass cluster is the single peak on the far right of these plots; it appears around the time of the peak in the SFR.

\begin{deluxetable}{ccccc}
\tabletypesize{\scriptsize}
\tablecaption{Mass Function Slopes at Fiducial Times \label{table:slopes}}
\tablewidth{0pt}
\tablehead{Time (Myr)&Slope\footnotemark&intercept}
\startdata  
  75&$   -0.94 \pm   0.65 $&$   3.52 \pm   1.96$\\
  128 &$  -1.33\pm    0.47$&$    4.93\pm    1.42$\\
  169  &$ -0.87 \pm   0.10$&$    4.65\pm    0.39$\\
  211 &$  -0.86 \pm   0.11$&$    4.73\pm    0.45$\\
  277 &$  -0.94 \pm   0.11$&$    5.19\pm    0.43$\\
\enddata
\footnotetext{These are slopes $\alpha$ on a $\log-\log$ plot where $-1$ is the conventional result. Subtract 1 to get the slope of the power law for the mass function itself, $n(M)\propto M^{\alpha-1}$.}
\end{deluxetable}

\begin{figure}
\includegraphics[width=\columnwidth]{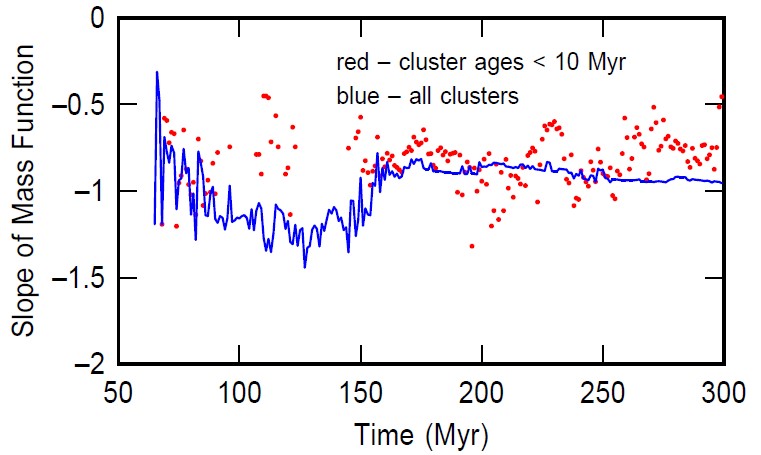}
\caption{The logarithmic slopes $\alpha$ of the mass functions for all clusters (blue) and young clusters (red), are shown versus time. They hover around $\alpha=-0.8$ to $-1$.  The mass function itself, $n(M)dM$, has a slope that is more negative by 1. } 
\label{mf_slopes}
\centering
\end{figure}

Figure \ref{mf_slopes} shows the time dependence of the slope of the mass function for all clusters (blue) and clusters younger than 10 Myr (red). The slope has a lot of scatter at first because there are not many clusters, but after the burst of star formation at $\sim150$ Myr, it settles down. The slope hovers around $\alpha\sim-0.8$ to $-1$ for all times and for both young clusters and all clusters. This consistency implies that the slope is a product of star formation. It also implies that highly chaotic gas motions during the galaxy collision and cluster feedback do not alter the property of the interstellar medium that determines the mass function. This is consistent with the idea that gas density structure from turbulence contributes to the universal mass function of star clusters. A similar result was shown in \cite{lahen22}, who also account for observational effects such as crowding and source confusion.

\begin{figure*}[hbt!]
\includegraphics[width=18cm]{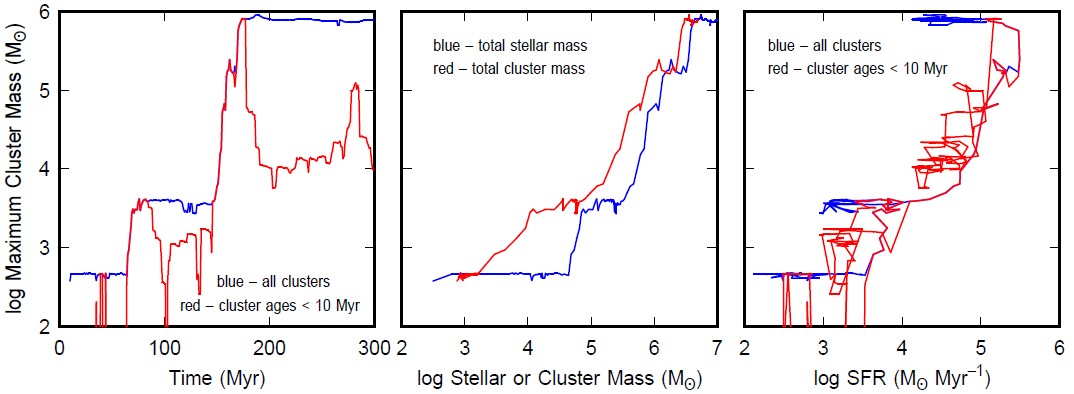}
\caption{The maximum cluster mass is shown versus time (left), total stellar or cluster mass (middle), and SFR (right). On the left and right, the blue curve is for all clusters and the red curve is for clusters younger than 10 Myr. The maximum mass of a young cluster follows the SFR even as it varies. The maximum mass for all clusters does not follow the SFR as well because massive clusters that formed in previous SFR bursts are still around when the SFR has dropped (this is why the blue horizontal segments appear in the left and right-hand plots). }
\label{max_mass}
\centering
\end{figure*}

\begin{figure}[b!]
\includegraphics[width=\columnwidth]{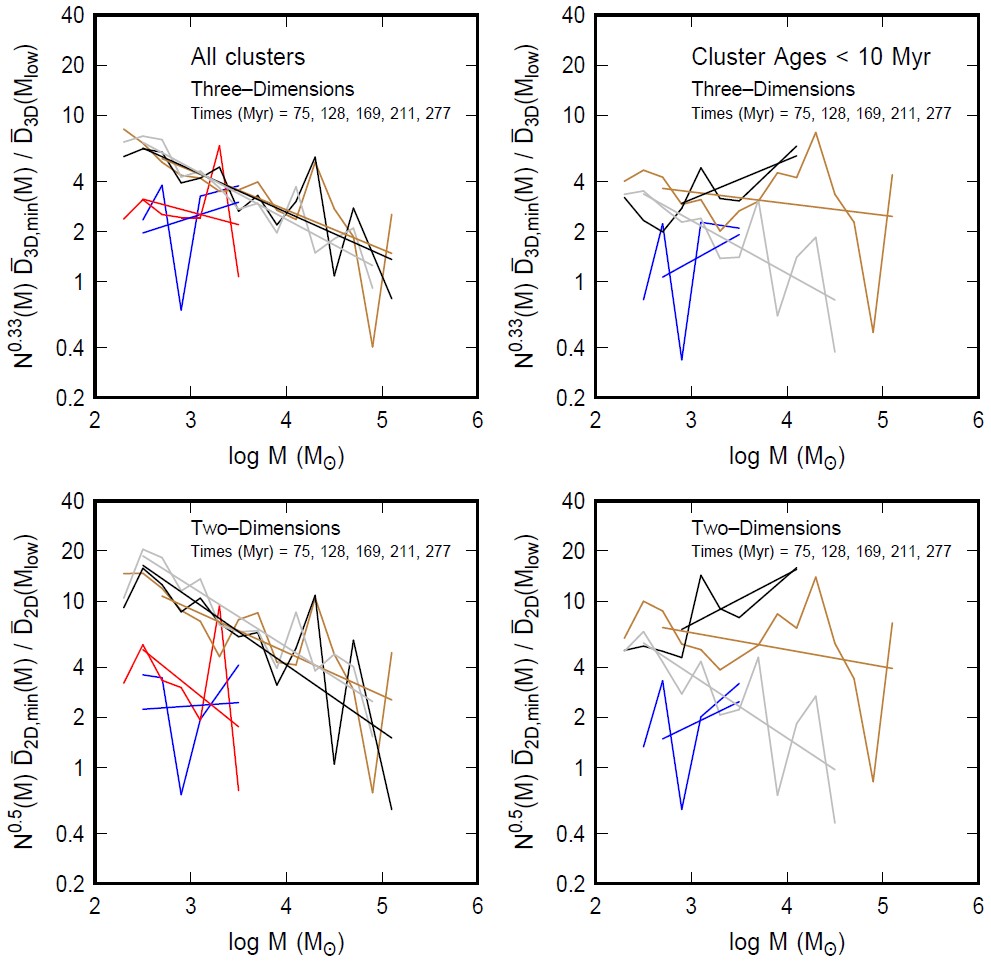}
\caption{The relative average minimum separation (${\cal R}(M)$ in eq. \ref{eq:R}) between equal-mass clusters is plotted as a function of cluster mass for mass intervals of 0.2 dex. Three-dimensional separations are in the top two panels and two-dimensional (projected) separations are in the bottom, while all clusters are shown on the left and young clusters are on the right. In all cases, when there are enough clusters to cover a wide mass range (fiducial times of 169, 211 and 277 Myr), the relative average minimum separation decreases with mass, indicating that for all cluster masses, the more massive clusters are closer to each other (at average minimum separation ${\bar D}_{\rm min}(M)$) than the average separation at their mass (which is proportional to $N^{-1/p}$ for $p=2$ and $p=3$ in two and three dimensions). The colors correspond to the times, as in Figure \ref{mf}.}
\label{ramps}
\centering
\end{figure}

\begin{figure}[b]
\includegraphics[width=8cm]{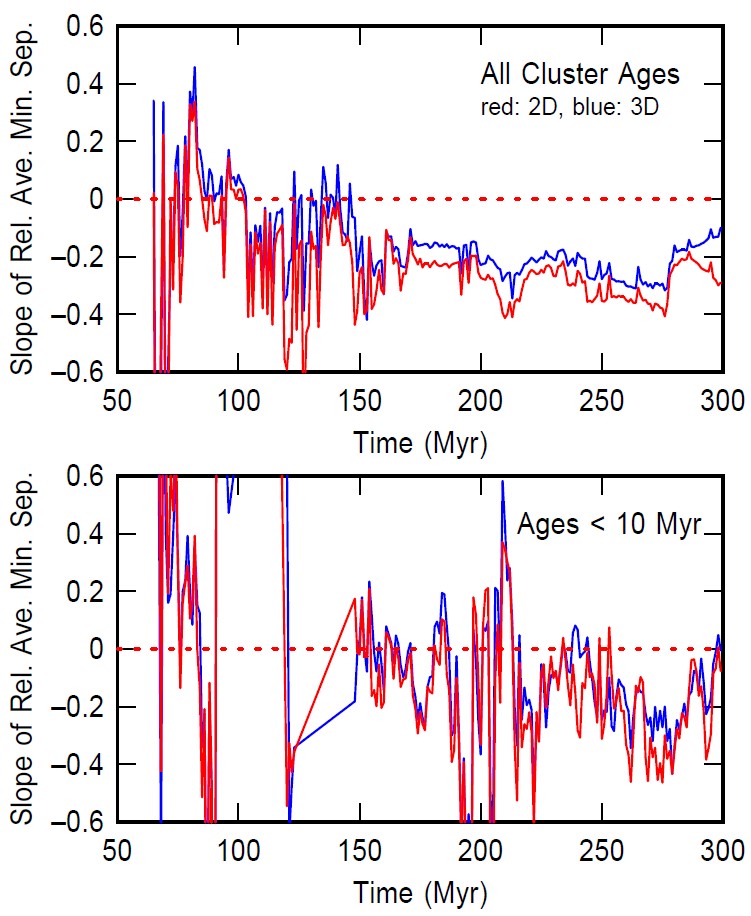}
\caption{The logarithmic slope of the power-law relation between the relative average minimum separation and the cluster mass, $d\log{\cal R}(M)/d\log M$, is shown versus time for all clusters (top) and young clusters (bottom) and for projected, 2D (red) and 3D (blue) separations. The slope hovers around $-0.2$.}
\label{ramps_slope}
\centering
\end{figure}

For a logarithmic mass function slope of $\alpha=-1$, the largest cluster mass is proportional to the number of clusters. Figure \ref{max_mass} shows the maximum cluster mass versus time in the left-hand panel, total stellar and cluster mass in the middle panel, and SFR in the right-hand panel. In the left and right-hand panels, the blue curves are for all cluster ages and the red curves are for clusters younger than 10 Myr. The maximum mass of a young cluster follows the SFR even as the SFR increases and decreases. The maximum mass for all ages does not follow the varying SFR because this mass does not decrease when the SFR drops.

\section{Relative Average Minimum Separation and Projected Separation}
\label{rampsection}
Clusters that are uniformly distributed in a volume $V$ have an average separation equal to the inverse cube-root of the density. This is true for all masses, which means that clusters in the mass range $M$ to $M+dM$ have an average separation equal to $(N(M)dM/V)^{-1/3}$ for mass function $N(M)$, equal to the number of clusters per unit mass. Clusters in a particular mass range that have a smaller average separation than expected from their number are systematically overdense. This property can be measured by $N^{1/3}{\bar D}_{\rm 3D,min}(M)$ for average {\it minimum} separation ${\bar D}_{\rm 3D,min}$ between all pairs of clusters of mass $M$ to $M+dM$. The average minimum separation is the average over all clusters (in that mass range) of the distances to the nearest cluster (in the same mass range). This quantity would be independent of mass if all clusters were randomly distributed in the volume, but it will decrease with increasing mass if massive clusters are more concentrated than low-mass clusters.  Because galaxies vary in their numbers of detectable clusters, it is useful to normalize this quantity to the average minimum separation between all clusters at the lowest mass in the power-law part of the cluster mass function, $M_{\rm low}$. Thus we use the quantity
\begin{equation}
{\cal R}(M)=N(M)^{1/3}{\bar D}_{\rm 3D,min}(M)/{\bar D}_{\rm 3D,min}(M_{\rm min})
\label{eq:R}
\end{equation}
where the 3D represents the 3D spatial distribution of clusters. A similar quantity can be defined for the projected, 2D distribution, or the deprojected 2D distribution (and with a power of $N$ equal to $1/2$). For the simulation, we consider both 3D and 2D distributions. In \cite{elmegreen20} we used the deprojected 2D distribution for observed galaxies, and referred to this quantity as the Relative Average Minimum Projected Separation (RAMPS). 

\begin{figure}
\includegraphics[width=\columnwidth]{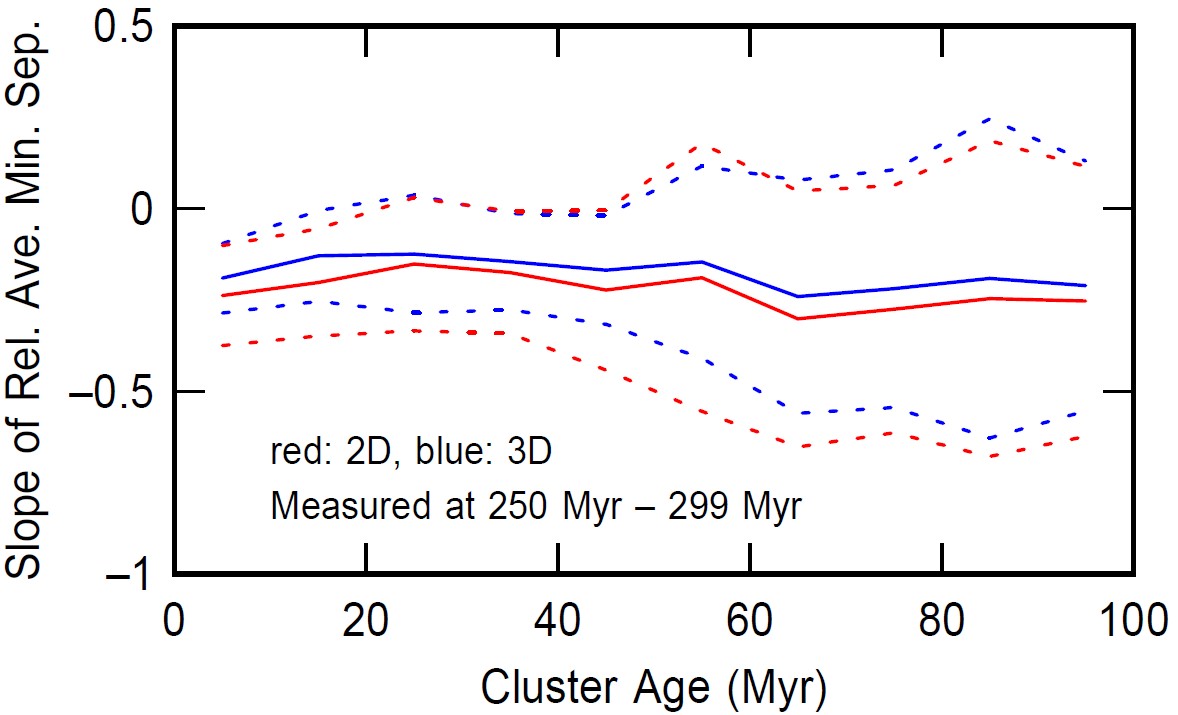}
\caption{The logarithmic slope of the power-law relation between the relative average minimum separation and the cluster mass, $d\log{\cal R}(M)/d\log M$, is shown versus cluster age for the time interval between 250 Myr and 299 Myr. Two-dimensional and three-dimensional cluster separations are shown in red and blue.  }
\label{slope_age}
\centering
\end{figure}

\begin{figure}[b]
\includegraphics[width=\columnwidth]{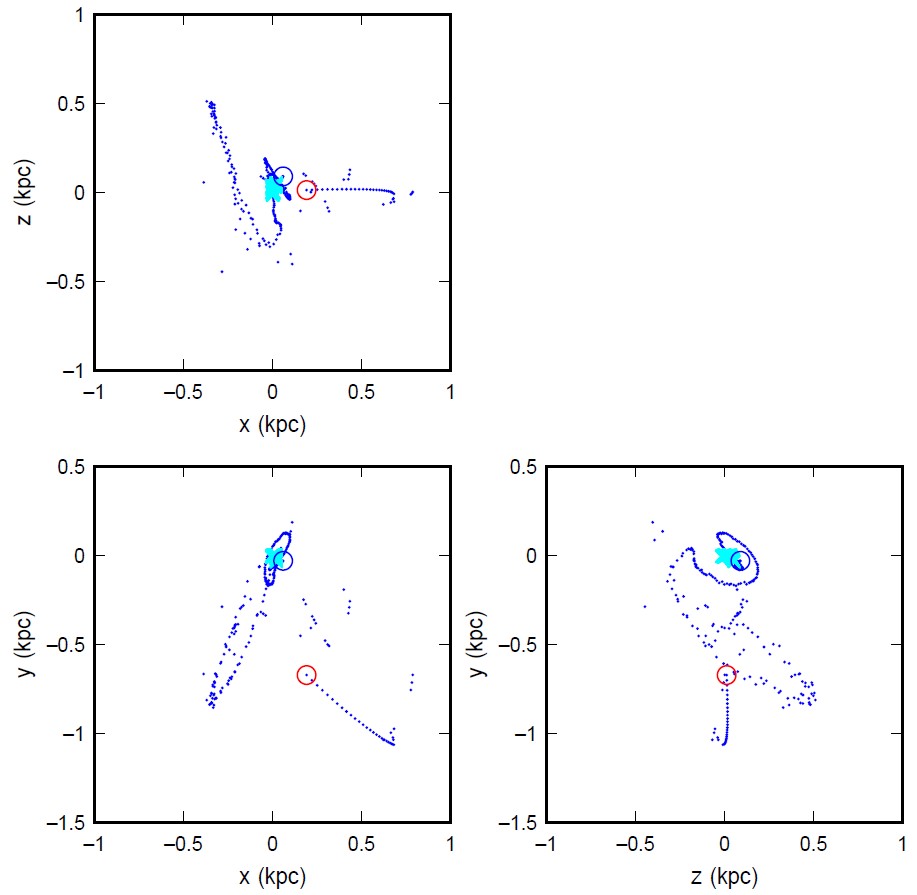}
\caption{The motion of the center of cluster positions is shown by a dot at each Myr interval. The beginning point is denoted by a red circle and the ending point by a blue circle. The center of mass for the dark matter moves a very small amount, as indicated by the overlapping cyan crosses. The three panels show three perspectives. The $(x,y)$ perspective in the lower left is used for the 2D calculations in this paper. }
\label{xyz}
\centering
\end{figure}

Figure \ref{ramps} shows ${\cal R}(M)$ for 3D (top) and 2D (bottom) distributions, and for all clusters (left) and clusters younger than 10 Myr (right). In each panel, ${\cal R}(M)$ is given at the 5 fiducial times with the same color code as in Figure \ref{mf}.  ${\cal R}(M)$ decreases with mass once there are enough clusters to obtain a meaningful slope, which is starting with the main starburst and for times of 169 Myr or later in the figure. Because both 3D and 2D curves have their minimum average separations ${\bar D}$ normalized at the lowest masses, their absolute values in these plots are proportional to $N(M)$ to the corresponding powers $1/3$ or $1/2$, making the 2D curves higher. The absolute value also depends on the mass interval used for the cluster counts, which is logarithmic in the figure, $d\log M=0.2$.

\begin{figure}[t]
\includegraphics[width=6cm]{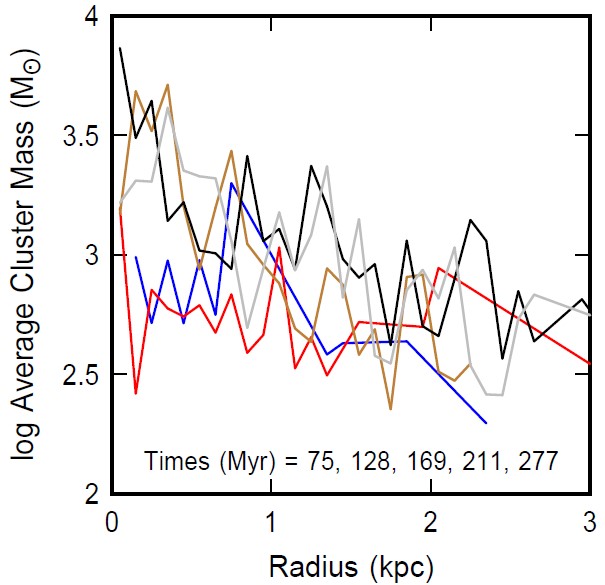}
\caption{The average cluster mass is shown as a function of distance to the center of the cluster distribution, which is the moving position mapped in Fig. \ref{xyz}. The five fiducial times are indicated by colors as in Fig. \ref{mf}. }.
\label{radial_gradient}
\centering
\end{figure}

Figure \ref{ramps_slope} shows the logarithmic slope $d\log{\cal R}(M)/d\log M$ versus time for 3D (blue) and 2D (red) distributions and for all clusters (top) and young clusters (bottom). Once there are enough clusters to have a reasonable slope, this slope is clearly negative even for young clusters. The 2D slope is slightly steeper than the 3D slope for the same reason that the 2D function ${\cal R}(M)$ is larger than the 3D function, namely the larger value of $N(M_{\rm min})^{1/2}$ than $N(M_{\rm min})^{1/3}$. 

\begin{figure}[b]
\includegraphics[width=6cm]{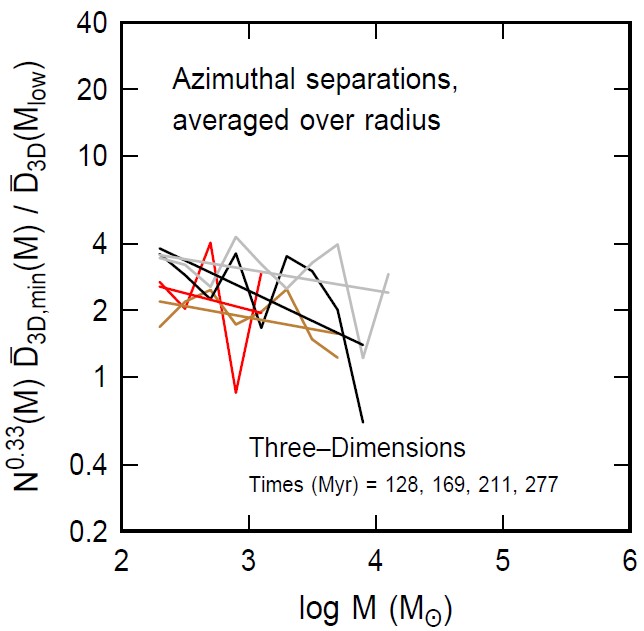}
\caption{The relative average minimum separation is shown as a function of cluster mass, as in Figure \ref{ramps}, but now considering only cluster pairs at the same distance from the center of the cluster distribution, to remove the radial gradient of cluster mass shown in Fig. \ref{radial_gradient}. The slope is still slightly negative. }
\label{ramps_azimuthal}
\centering
\end{figure}

The early appearance of negative slopes in Figure \ref{ramps_slope} implies that cluster mass segregation is a result of star formation, like the mass function. Figure \ref{slope_age} confirms this by plotting the average slope versus cluster age for a simulation time interval near the end, between 250 Myr and 299 Myr, and for cluster ages in intervals of 10 Myr. The dotted lines signify the uncertainties in the slopes. The uncertainty gets larger for larger ages because there are fewer old clusters at this time. Even the youngest clusters are segregated by mass. 

\begin{figure}[t]
\includegraphics[width=\columnwidth]{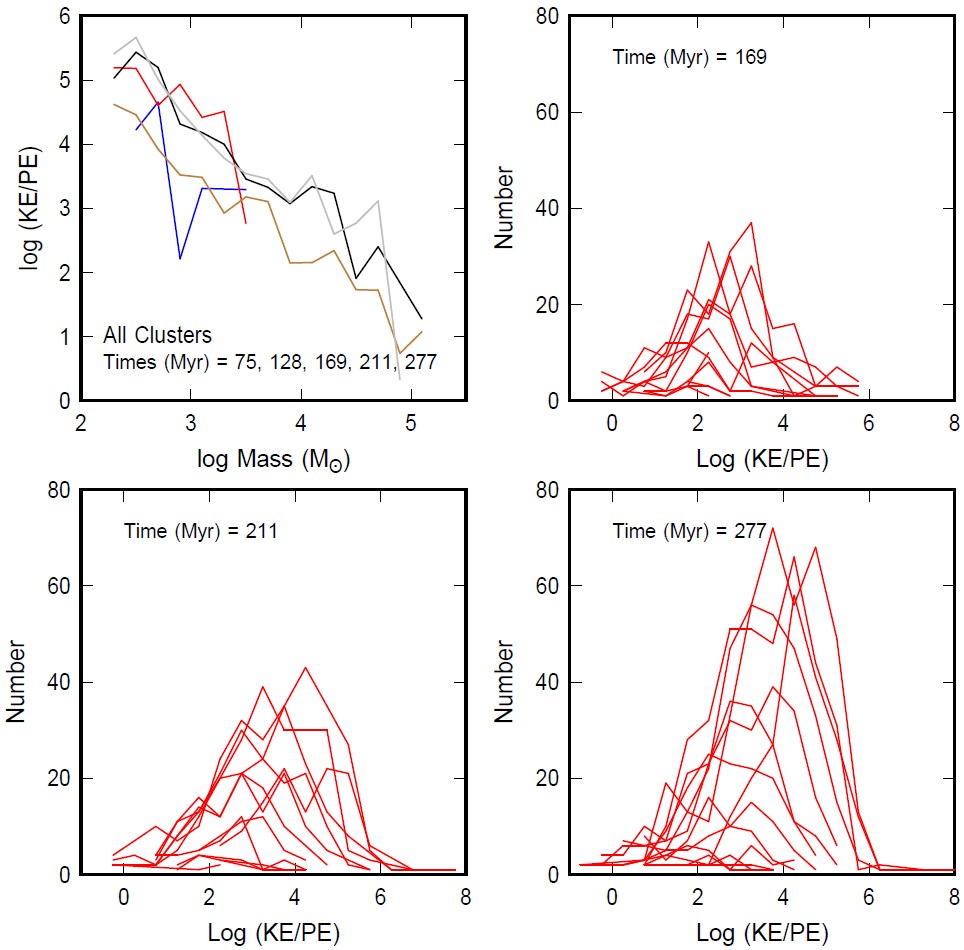}
\caption{(top left) The ratio of the kinetic energy to the potential energy is shown as a function of cluster mass for pairs of nearest-neighbor clusters with similar masses, as divided into 0.2 dex mass intervals. The ratio is generally high, $>>10$, but the distribution functions of this ratio, shown for the fiducial times in the other panels, indicate a few cluster pairs with low ratios, $\log(KE/PE)<0$. Such low-ratio clusters are prone to coalescence.}
\label{kepe}
\centering
\end{figure}

The global distribution of clusters is highly irregular in this interacting pair with the center of the cluster coordinates moving nearly a kpc in distance while the interaction is taking place. Figure \ref{xyz} shows the center at each Myr interval.  This irregularity is the result of the galaxy merger and is not a general feature of local dwarf galaxies.  It is useful for the present study because we can view the cluster distribution  independently of the galactic radial distribution, which cannot be done for symmetric galaxies. Nevertheless, there is an example of massive clusters together in a compact region, perhaps showing mass segregation relative to other clusters, in a lensed dwarf at redshift $\sim10.2$ \citep{adamo24}.

The segregation of massive clusters illustrated by the negative slopes in Figures \ref{ramps} and \ref{ramps_slope} corresponds to a centralization within the entire distribution of clusters, which, as seen in Figure \ref{xyz}, is distinct from a centralization in the galaxy-pair relative to the dark matter center of mass. Figure \ref{radial_gradient} shows the average cluster mass versus distance from the instantaneous center of the cluster distribution, i.e., from the points in Figure \ref{xyz}. There is a decreasing trend showing a segregation of more massive clusters toward this center. To check whether the entire segregation effect shown by ${\cal R}(M)$ is the result of this radial mass dependence, Figure \ref{ramps_azimuthal} shows a modified ${\cal R}(M)$ function where only clusters considered within the same radial interval are used to search for near neighbors. That is, the radial range is divided into 500 pc bins (0-500pc, 500pc-1000pc, etc.) and near neighbors for each cluster are searched entirely within its bin, and also within a maximum distance proportional to the bin radius, namely within 250pc for the 0-500 pc bin, within 750 pc for the 500-1000pc bin, etc., to prevent cluster pairs from opposite sides of the galaxy. The result in Figure \ref{ramps_azimuthal} shows the trend. The fits have logarithmic slopes of $-0.15\pm1.09$, $-0.10\pm0.16$, $-0.27\pm0.25$, $-0.09\pm0.15$ for the four times, in order.

\section{Cluster Coalescence}
\label{coalesce}
The clusters in this galaxy collision are moving rapidly so relatively few are gravitationally attracted to each other strongly enough to coalesce. To determine the fraction of close cluster pairs that might coalesce, we determined the ratio of the kinetic energy, $KE$, to the potential energy, $PE$, for  nearest-neighbor pairs, 
\begin{equation}
{{KE}\over{PE}} = {{0.5\mu v^2 R}\over{GM_1M_2}},
\end{equation}
where $\mu=M_1M_2/(M_1+M_2)$ is the reduced mass for cluster masses $M_1$ and $M_2$, $R$ is the cluster separation, and $v$ is the relative velocity given by, $v^2=(v_{\rm x1}-v_{\rm x2})^2+(v_{\rm y1}-v_{\rm y2})^2+(v_{\rm z1}-v_{\rm z2})^2$.  We first do this for 0.2 dex intervals of mass, where the nearest neighbors are in the same mass interval, and then we consider all nearest-neighbor pairs, regardless of mass. 

Figure \ref{kepe} shows the average value of the log of the ratio of KE to PE as a function of mass for the first case, where each cluster in the pair has about the same mass. The five fiducial times are plotted separately with the same color code as in previous figures. The decreasing trend indicates that low-mass clusters typically move too fast to deflect each other gravitationally ($KE/PE\sim10^4$), but high mass clusters come close to being mutually gravitating ($KE/PE\sim10$). Because these are averages and there is a dispersion of values for each average, we plot in the other panels the distribution functions of $KE/PE$ for the last three fiducial times, which have the most clusters. Each curve is a different mass interval (curves for lower masses have higher $KE/PE$). We note that several clusters have $KE/PE\sim1$, at which point these clusters should strongly attract each other, deflect in their motion, and possibly coalesce.
\begin{figure}[t!]
\includegraphics[width=\columnwidth]{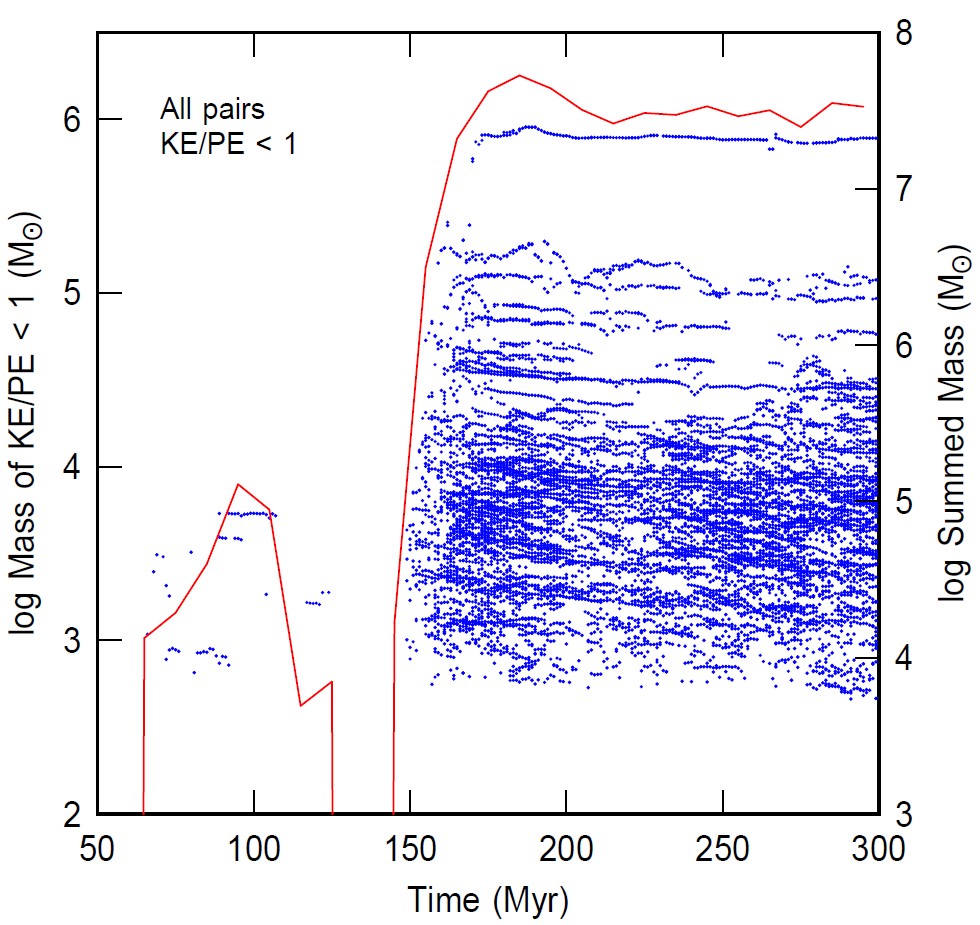}
\caption{The summed mass of a cluster pair is shown versus time in the simulation for all cluster pairs that are nearest neighbors regardless of mass, and which also have a ratio of relative kinetic energy to potential energy less than 1. Such near-neighbor pairs are likely to coalesce. Most of the thin horizontal streaks are single cluster pairs that have $KE/PE<1$ for the entire time of the streak; the clusters coalesce at the right-hand end of the streak. Compared to all nearest-neighbor cluster pairs, those with $KE/PE<1$ represent a fraction of 8.3\%. The red curve represents the total cluster mass with $KE/PE<1$ as a function of time, using the right-hand axis. }
\label{kepe_dist}
\centering
\end{figure}

To track such incidences in more detail, we plot in Figure \ref{kepe_dist} the summed mass of clusters in a pair versus the time for all cluster pairs that have $KE/PE<1$, regardless of mass. Each point in the plot indicates a cluster pair that is likely to coalesce. The points produce thin horizontal streaks in this plot when cluster pairs maintain $KE/PE<1$ over a period of time. The last time in a streak corresponds either to coalescence of the clusters or disruption of the pair. The red curve in the figure is the summed mass of all clusters with $KE/PE<1$ as a function of time, using the right-hand ordinate for the scale. 
The number of nearest-cluster pairs with $KE/PE<1$ in the figure is 8821, which is 8.3\% of the total number of nearest-cluster pairs at any $KE/PE$ value in all timesteps. If $KE/PE<1$ is an indication of eventual coalescence, then about this fraction of clusters coalesce.

\begin{figure}
\includegraphics[width=\columnwidth]{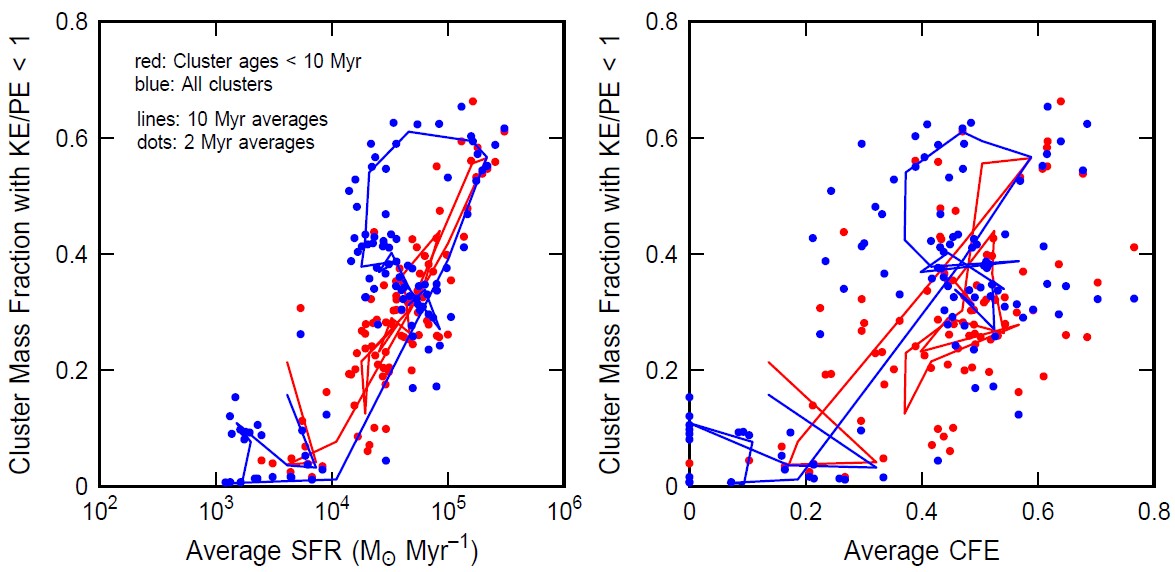}
\caption{The fractional mass of all clusters that have nearest neighbors with $KE/PE<1$ is shown versus the average star formation rate and the average cluster formation efficiency. The blue points and lines consider all clusters, and the red points and lines are for clusters only younger than 10 Myr. The lines connect averages in both the abscissae and ordinates in 10 Myr intervals, and the dots are for 2 Myr averages, showing shorter-time variations. }
\label{kepe_sfr_cfe}
\centering
\end{figure}

The fractional mass of cluster pairs with $KE/PE<1$ is shown versus the SFR and CFE in Figure \ref{kepe_sfr_cfe}. The fractional mass of all cluster pairs, regardless of cluster age, is shown in blue, and the fractional mass for just the youngest clusters, where both members of the pair have ages less than 10 Myr, is shown in red. The lines in the plots connect the values calculated with 10 Myr averages, and the dots are for 2 Myr averages, showing the temporal scatter around the lines. The mass fraction susceptible to coalescence ($KE/PE<1$) is essentially zero for SFRs less than $\sim10^4\;M_\odot$ Myr$^{-1}$ and then it rises rapidly with SFR, reaching a peak mass fraction of about 0.6 at $10^5\;M_\odot$ Myr$^{-1}$. For the young clusters (red dots), the fractional mass tracks the SFR above this limit. Note that the mass fraction of $KE/PE<1$ cluster pairs is much larger than the number fraction of  $KE/PE<1$ cluster pairs, which was discussed in the previous paragraph. This is because massive clusters are more likely to coalesce than low-mass clusters (Fig. \ref{kepe}). 

Figure \ref{kepe_sfr_cfe}(left) resembles Figure \ref{sfr}(right) in the threshold behavior for SFR, which is why figure \ref{kepe_sfr_cfe}(right) has a nearly linear relation, although with a lot of temporal scatter. A plausible reason for the good correlation between the $KE/PE<1$ mass fraction and the CFE is that higher SFRs produce much higher densities of young clusters, first because the gas densities in the regions are higher, second because the density of young stars is higher with both the higher density and the higher SFR, and third because the cluster formation efficiency is higher with the higher SFR.  As a result, clusters are closer to each other and more prone to coalescence at higher CFE. Conversely, the higher CFE could be the result of increased cluster coalescence at higher SFR if non-clustered stars are brought into the coalescence along with clustered stars. 

\section{Discussion}
\label{discussion}

The details of how the hierarchical assembly of star clusters proceeds may have a profound impact on the internal evolution of the clusters. The frequency of stellar interactions in the central regions of star clusters depends on the evolution of the density structure of the clusters \citep{fujii2013}. For instance, the mass-growth of stars and black holes through collisions in hierarchical star-forming regions can be boosted compared to monolithic star clusters \citep{rantala2024}. This mass-growth impacts the chemical and radiation output of the stellar population in the clusters. Hydrodynamical simulations that include chemical enrichment through single very massive stars \citep{lahen2024} call for additional feedback channels to explain the large number of light-element enriched stars observed in globular clusters \citep{bastian2018}. Enhanced production of active, very massive stars in hierarchically assembled massive star clusters could contribute to increased chemical variations in those clusters.

The radial segregation of massive clusters could have implications for the assembly of nuclear star clusters. Recent observational surveys have found dwarf galaxies to host nuclear clusters that possibly formed through accretion of globular clusters instead of in-situ star formation \citep{fahrion21,fahrion22}. In the absence of a dark matter cusp or central massive black hole in our simulated dwarf galaxies, the massive clusters that form in the central starburst end up oscillating on almost radial orbits within 1 kpc of the galactic centre. With a dark matter cusp, dynamical friction should cause the massive star clusters to sink further inward in the galaxy \citep{meadows20,modak23} and collide with other clusters formed in the galaxy or brought in through successive galaxy mergers. A cosmological simulation with more complex assembly histories for dwarf galaxies (see e.g. \citealt{gutcke22}) would be needed to follow the possible coalescence and growth of nuclear star clusters.

\section{Conclusions}\label{conclusions}
Star clusters that form in the dwarf galaxy collision simulated by \cite{lahen20} have the usual power-law mass function with logarithmic slope $\alpha\sim-0.8$ to $-1$ from the time of their birth, with a maximum cluster mass that scales with the instantaneous SFR. The clusters also segregate in mass from the time of their birth, with more massive clusters centrally concentrated relative to each other than less massive clusters. Near-neighbor clusters are generally moving too fast to deflect by gravitational forces, but $\sim8$\% of them become significantly attracted to each other during the simulation, with a relative kinetic energy less than their potential energy. These 8\% could coalesce. They represent up to 60\% of the cluster mass at the peak of the SFR.

The implications of these results are that the mass distribution function and the spatial segregation of star clusters according to their masses are products of their birth that resist disruption for at least several hundred Myr. These properties are continuously present as new clusters appear, suggesting that the conditions for cluster formation, such as the intrinsic structure and motion of interstellar gas, are always present, regardless of how chaotic these conditions get from the galaxy collision and stellar feedback.  The preferred centralized location of  massive clusters demonstrates either an environmental effect for cluster mass or a migration of massive clusters soon after they form. Environmental effects add physical processes to the determination of cluster mass, in addition to purely stochastic sampling of the mass function \citep{pflamm13}. Hierarchical cluster assembly has further implications for the formation of supermassive stars in cluster cores, while cluster mass segregation could promote the formation of nuclear star clusters. 

\section{Acknowledgements}
This work was initiated in the MIAPbP workshop ``Star-forming clumps and clustered starbursts across cosmic time'' from October 4 to 28, 2022. 
This research was supported by the Munich Institute for Astro-, Particle
and BioPhysics (MIAPbP) which is funded by the Deutsche
Forschungsgemeinschaft (DFG, German Research Foundation) under Germany´s
Excellence Strategy – EXC-2094 – 390783311. NL acknowledges the computing time granted by the LRZ (Leibniz-Rechenzentrum) on SuperMUC-NG (project number pn49qi) and The Max Planck Computing and Data Facility (MPCDF) in Germany and CSC -- IT Center for Science Ltd. in Finland.

\end{document}